\documentclass[twocolumn,printnumbers,amsmath,amssymb,prb]{revtex4}
\usepackage{graphicx}
\usepackage{color}

\makeatletter
\renewcommand{\section}{\@startsection{section}{1}{0mm}
  {-\baselineskip}{0.5\baselineskip}{\bf\leftline}}
\makeatother

\makeatletter
\renewcommand{\subsection}{\@startsection{subsection}{1}{0mm}
  {-\baselineskip}{0.5\baselineskip}{\bf\leftline}}
\makeatother

\begin{document}

\title{Energy Transport in Glasses}

\author{Elijah Flenner$^1$} 
\author{Lijin Wang$^2$}
\author{Grzegorz Szamel$^1$} 

\affiliation{$^1$Department of Chemistry, Colorado State University, Fort Collins, Colorado 80523, USA \\
$^2$School of Physics and Materials Science, Anhui University, Hefei 230601, P. R. China. }

\begin{abstract}
The temperature dependence of the
thermal conductivity is linked to the nature of the energy transport at a frequency $\omega$, which is quantified by
thermal diffusivity $d(\omega)$. Here we study $d(\omega)$ for a poorly annealed glass and a highly
stable glass prepared using the swap Monte Carlo algorithm. To calculate $d(\omega)$, we excite wave packets and find that
the energy moves diffusively for high frequencies up to a maximum frequency, beyond which the
energy stays localized. At intermediate
frequencies, we find a linear increase of the square of the width of the wave packet with time, which allows for
a robust calculation of $d(\omega)$, but the wave packet is no longer well described by a Gaussian
as for high frequencies. In this intermediate regime, there is a transition from a nearly frequency independent
thermal diffusivity at high frequencies to $d(\omega) \sim \omega^{-4}$  at low frequencies.
For low frequencies the sound waves are responsible for energy transport and the energy moves ballistically. The
low frequency behavior can be predicted using sound attenuation coefficients.
\end{abstract}

\maketitle

The thermal conductivity of amorphous solids is vastly different than that of their crystalline counterparts.
The existence of several common features in the temperature dependence of the thermal conductivity
of amorphous solids indicates a common origin \cite{Eucken1911,Kittel1949,Zeller1971,Zaitlin1975,Anderson1972,Phillips1972,Pohl2002,Stephens1973}.
At temperatures below $\sim 1$K the thermal conductivity grows as
$T^2$ compared to $T^3$ growth for crystalline solids. This
quadratic growth of the thermal conductivity with temperature is generally attributed to two-level
tunneling states \cite{Zeller1971,Anderson1972,Pohl2002,Lubchenko2001,Lubchenko2003,Lubchenko2018},
although alternative explanations exist \cite{Leggett2013,Yu1991,Burton2016}.
Around $T \approx 10K$ a plateau develops in the thermal conductivity and
there is a nearly linear rise in the thermal conductivity after the plateau.

The temperature dependence of the thermal conductivity $\kappa$ can be analyzed in terms of
frequency dependent thermal diffusivity $d(\omega)$,
which quantifies how fast a wave packet, narrowly peaked around a frequency $\omega$,
propagates \cite{Allen1989,Allen1993,Xu2009,Vitelli2010}.
At low temperatures, only the low frequency modes are excited, and only the
low frequency thermal diffusivity significantly contributes to the thermal conductivity.
The most prevalent theories attribute the low frequency thermal diffusivity to two-level states, which provide the dominant
contribution below 1K, and to thermal transport due to sound waves \cite{Bucenau1992,Parshin1994}.
By considering the sound waves
as a phonon gas, Debye argued that there is a contribution to $d(\omega)$ given by $v(\omega) \ell(\omega)/3$ where $v(\omega)$
is the speed of sound and $\ell(\omega)$ is the mean free path \cite{Debye1912}.
It is often assumed, and confirmed in recent simulations, that
sound attenuation obeys Rayleigh scaling, and thus
the contribution due to sound waves behaves as $d_s(\omega) \sim \omega^{-4}$ \cite{Lerner2016,Wang2019,Mizuno2018}.
Several researchers demonstrated that the thermal conductivity can be accurately
described for temperatures at and below the low temperature plateau by combining the contributions to
$d(\omega)$ due to two level systems and due to sound waves
\cite{Bucenau1992,Parshin1994,Feldman1993,Shirmacher2006}.

At room temperature, where all vibrational modes are excited,
the average mean free path is on the order of the interatomic spacing \cite{Feldman1993,Cahill1989,Beltukov2018,Kittel1996},
which implies that the high frequency
excitations are strongly damped and can no longer be described as propagating sound
waves. This strong damping is consistent with simulations, which show that energy transport does not have
the  low frequency ballistic character associated with sound waves\cite{Beltukov2018,Beltukov2016,Sheng1991}.
Instead, for high frequencies the energy transport is diffusive.
Xu \textit{et al.}\cite{Xu2009} showed that for systems of jammed spheres the thermal diffusivity is constant above a characteristic
frequency $\omega_d$ up to a maximum frequency $\omega_c$ where it quickly drops
to zero. The crossover frequency $\omega_d$ goes
to zero as the unjamming transition is approached. This constant diffusivity can explain the linear
increase of the thermal conductivity above the plateau. For high frequencies, the thermal diffusivity goes to zero and
the excitations are localized \cite{Xu2009,Vitelli2010,Beltukov2018,Beltukov2016,Sheng1991}.

Identification of these three regimes for $d(\omega)$ motivated Allen \textit{et al.}\cite{Allen1989,Allen1999} to characterize
the vibrational modes in terms of propagons, diffusons, and locans. They determined that for amorphous silica
97\% of modes are diffusons, which implies that diffusive transport is the  dominant contribution to
the thermal conductivity for temperatures above the plateau.

Few simulations have studied the full range
of diffusivity from the low-frequency sound wave dominated regime to the high-frequency plateau \cite{Beltukov2018,Beltukov2016}.
The method of Allen and Feldman \cite{Allen1989, Allen1993} is currently restricted to high frequencies since one needs to
diagonalize the Hessian, which restricts one to small systems. Here, we use an alternative method \cite{Beltukov2018,Beltukov2016,Sheng1991}
to study the full range
of frequencies, the crossover between high and low frequency, and the connection
between the diffusivity and phonon attenuation.
Additionally, to our knowledge, the energy diffusivity has not been studied as a function of the
glass stability. Since the vibrational properties and the attenuation of sound waves
change dramatically with stability \cite{Wang2019,Wang2019a,Lerner2019JCP},
the energy diffusivity would also be expected to change with
stability. Here, we compare the energy diffusivity over a broad range of frequencies for
a poorly annealed glass and a glass whose stability is similar to that of stable laboratory glasses.

To calculate the thermal diffusivity we excite a narrow, in frequency and space,
wave packet at the center of a simulation cell and examine
energy transport in the harmonic approximation. For diffusive energy transport, the center
of the wave packet remains stationary and the square of its width increases as $2 d(\omega) t$
in each direction. We find that this method results in the same diffusivity as the method of
Allen and Feldman \cite{Allen1989, Allen1993} using the eigenvectors and eigenvalues of the Hessian. Since we
excite a wave packet that can only propagate in one direction, we simulate elongated systems.
These elongated systems allow us to extend the time scale for the energy transport calculation,
and thus examine energy diffusivity at low frequencies where the transport is ballistic.
Therefore, we can investigate the crossover from diffusive to ballistic energy transport.



\section{Simulations}

We create glasses by quenching a polydisperse model glass former equilibrated at a
parent temperature $T_p$ to its inherent structure using
the fast inertia relaxation engine minimization algorithm \cite{fire}.
The interaction between two particles $n$ and $m$ is given by
\begin{equation}
V(r_{nm}) = \epsilon \left( \frac{\sigma_{nm}}{r_{nm}} \right)^{12} + v(r_{nm})
\end{equation}
when $r_{nm} = | \mathbf{r}_n - \mathbf{r}_m | < 1.25 \sigma_{nm}$ and zero otherwise.
The continuity of $V(r_{nm})$ is ensured up to the second derivative at the cutoff by setting
$v(r_{nm}) = c_o + c_2 (r_{nm}/\sigma_{nm})^2 + c_4 (r_{nm}/\sigma_{nm})^4$. The probability
that a particle has a diameter $\sigma$ is given by $P(\sigma) = A/\sigma^3$ where
$\sigma \in [0.73,1.63]$, and we use a non-additive mixing rule $\sigma_{nm} = 0.5(\sigma_n + \sigma_m)(1-0.2|\sigma_n - \sigma_m|)$.
To equilibrate the systems at $T_p =0.2$ and $T_p =0.062$ we use the Monte Carlo swap algorithm \cite{Berthier2016,Ninarello2017}.
The higher parent temperature is approximately equal to the onset temperature for the slow dynamics and
the resulting inherent structure constitutes a poorly annealed glass. The lower parent temperature is lower than
the estimated experimentally equivalent glass transition temperature of $T_g \approx 0.072$ \cite{Ninarello2017}. The
inherent structure resulting from quenching the sample equilibrated at $T_p =0.062$ constitutes a very stable glass.
We present the results in reduced units where $\epsilon$ is the unit of energy and $\sqrt{M \sigma^2/\epsilon}$ is the unit
of time.  Each particle has the same mass $M$, which is our mass unit. We set  Boltzmann constant $k_B = 1$.

We equilibrated systems of $N=3\, 000$ and $N=48\, 000$ particles at a number density $\rho = N/V = 1.0$. Since at low frequencies
the energy moves ballistically at the speed of sound, we needed large systems. To this end we replicated the $N=3\, 000$ particle system
80 times in the x-direction to make a very elongated simulation box with $243\, 000$ particles. We replicated the $48\, 000$ particle
system two times to make a $144\, 000$ particle system. We have checked that there were no finite size effects.

To study the energy transport we excited a wave-packet centered at $x=0$. To this end we solved the harmonic equations of
motion
\begin{equation}
\label{eqmotion}
\ddot{\mathbf{u}}_n(t) = -\sum_{n=1}^N \mathbf{D}_{nm} \cdot \mathbf{u}_m(t) + \mathbf{f}_n(\phi,\omega,x,t),
\end{equation}
where $\mathbf{u}_n = \mathbf{r}_n - \mathbf{r}_n^0$, $\mathbf{r}_n^0$ is the inherent structure
position, $\mathbf{D}_{nm}$ is the dynamical matrix (Hessian).
The external force exciting the wave packet, $\mathbf{f}_n(\phi,\omega,x,t)$, is given by
\begin{eqnarray}
\mathbf{f}_n(\phi,\omega,x,t) &=& \mathbf{a}_\lambda \cos(\omega t + \phi) \nonumber \\
&& \times \exp\left[ -\frac{1}{2}\left(\frac{x}{\Delta x}\right)^2 - \frac{1}{2} \left( \frac{t}{\Delta t}\right)^2 \right].
\end{eqnarray}
We started the simulation at $t = -5\Delta t$ so that $\mathbf{f}_n(\phi,\omega,x,t) \approx 0$ and run until the excitation reaches
the end of the simulation box. Unless otherwise noted, we use $\Delta x = 0.5$ and $\Delta t = 10$. Since a
wave-packet of finite duration is a mixture of different frequencies, our frequency uncertainty
is $\Delta \omega \approx 1/(\Delta t) = 0.1$.

\section{Energy Transport Calculation}

We use an approach proposed by Beltukov \textit{et al.}\cite{Beltukov2018} and run two simultaneous simulations using the same system.
The simulations differ by the phase $\phi$ in the external force exciting the wave packet. For the first simulation
$\phi = 0$ and for the second simulation $\phi = \pi/2$.
Alternatively, one can run one simulation and divide the kinetic and potential energy into regions, but it is ambiguous
how to divide the potential energy between the two interacting particles. Beltukov \textit{et al.}'s approach removes that ambiguity.

The energy is converted from potential to kinetic at a rate given by $\omega$ and
the energy density can be defined as
\begin{equation}
E(\omega,x,t) = \frac{1}{2 l_y l_z} \sum_n \left[ (\dot{\mathbf{u}}_n^0)^2 + (\dot{\mathbf{u}}_n^{\pi/2})^2 \right] \delta(x - x_n),
\end{equation}
where $\dot{\mathbf{u}}_n^\phi$ is the velocity of particle $n$ in simulation with phase $\phi$ at a time $t$,
$l_y$ is the box length in the $y$ direction, and $l_z$ is the
box length in the $z$ direction. We study longitudinal excitations by setting $\mathbf{a}_L = (a,0,0)$ and transverse excitations by
setting $\mathbf{a}_T = (0,0,a)$. We also study random excitations that are described by
$\mathbf{a}_T = (a r_x, a r_y, a r_z)$ where $r_x$, $r_y$, and $r_z$ are
Gaussian distributed random numbers of unit variance. Since we are using the harmonic approximation,
the results are independent of the value of $a$. If energy transport is diffusive,
the thermal diffusivity $d(\omega)$ can be calculated by calculating
$\delta r^2(\omega,t) = \int dx x^2 E(\omega,x,t)/\int dx E(\omega,x,t)$
and fitting $\delta r^2(\omega,t) = 2 d(\omega) t + r_0^2$ for the range of times when $\delta r^2(\omega,t)$ is linear.  If the
energy transport is ballistic, then $\delta r^2(t) \sim t^2$.

An alternative approach to determine the thermal diffusivity is due to Allen and Feldman \cite{Allen1989,Allen1993}. This
approach was used in several simulations utilizing the harmonic approximation  \cite{Xu2009,Vitelli2010,Beltukov2018,Beltukov2016}.
Within Allen and Feldman's approach, the thermal diffusivity is determined from the eigenvalues and eigenvectors of the Hessian matrix.
This approach is time consuming and suffers from finite size effects \cite{Bouchbinder2018,Wang2019a}.
We used the method of Allen and Feldman and compared the resulting
thermal diffusivity with the thermal diffusivity obtained from Beltukov \textit{et al.}'s approach.

According to Allen and Feldman's approach, the thermal diffusivity can be calculated as follows
\begin{eqnarray}
\label{daf}
d_{AF}(\omega) &=& \frac{\pi}{12 M^2 \omega^2} \int_0^\infty d\omega^{\prime} D(\omega^\prime) \nonumber \\
&& \times \frac{(\omega + \omega^\prime)^2}{4 \omega \omega^\prime} \left| \mathbf{S}(\omega,\omega^\prime)\right|^2 \delta(\omega-\omega^\prime).
\end{eqnarray}
The heat-flux matrix elements are given by
\begin{equation}
\left|\mathbf{S}(\omega,\omega^\prime)\right|^2 =
\frac{\sum_{mn} \left|\mathbf{S}_{nm}\right|^2 \delta(\omega-\omega_m)\delta(\omega^\prime-\omega_n)}{D(\omega) D(\omega^\prime)},
\end{equation}
where the sum is over the vibrational modes. The matrix elements $\mathbf{S}_{mn}$ are
\begin{equation}
\mathbf{S}_{mn} = \sum_{i,j} (\mathbf{r}_i - \mathbf{r}_j) \mathbf{e}_{m,i} \cdot \mathbf{D}_{i,j} \cdot \mathbf{e}_{n,j},
\end{equation}
where $\mathbf{e}_{n,i}$ is the normalized eigenvector of the Hessian matrix.
For a finite system the delta function in
equation \ref{daf} is replaced by $g(\omega_m - \omega_n,\eta) = \eta/\{ \pi [(\omega_m-\omega_n)^2 + \eta^2]\}$
where we have set $\eta = 0.01$.

\section{Thermal Conductivity}
The thermal conductivity $\kappa$ can be expressed in terms of $d(\omega)$, the density of
states $D(\omega) = \sum_m \delta(\omega - \omega_m)$ and the heat capacity $C(\omega,T)$
using the following formula\cite{Sheng1991,Allen1993,John1983}
\begin{equation}
\label{kappa}
\kappa = \frac{1}{V} \int_0^{\infty} d\omega D(\omega) d(\omega) C(\omega,T).
\end{equation}
The heat capacity $C(\omega,T) = k_B (\beta \hbar \omega)^2 e^{\beta \hbar \omega}/(e^{\beta \hbar \omega} - 1)^2$,
where $\beta = 1/k_B T$,   $T$ is the temperature, and $\hbar$ is the reduced Planck constant.

Although according to the standard nomenclature $d(\omega)$ is referred to as the energy diffusivity, the energy
transport does not have to be diffusive, and it can arise from other mechanisms.
Here we calculate $d(\omega)$ within the classical harmonic approximation, and thus we ignore
anharmonic effects that are important to understanding the full temperature dependence of $d(\omega)$. The most important
neglected effect is proposed to be scattering due to two-level systems that is both quantum mechanical and anharmonic
\cite{Zeller1971,Anderson1972,Pohl2002}.  Our approach
marks a starting point and quantum mechanical and anharmonic effects are left for future work.

When there is more than one energy transport
mechanism, it is usually assumed that $d^{-1}(\omega) = \sum_n d_n^{-1}(\omega)$ where
$d_n(\omega)$ correspond to different energy transport mechanisms \cite{Bucenau1992,Parshin1994,Feldman1993,Shirmacher2006}.
At low frequencies, the dominant energy transport mechanism is sound waves
(in the harmonic approximation) and $d^{-1}(\omega) \approx d_L^{-1}(\omega) + d_T^{-1}(\omega)$
where $d_L(\omega)$ is the contribution due to longitudinal sound waves and $d_T^{-1}(\omega)$ is due to the transverse waves.
It is expected that the energy transport is dominated by the transverse waves,
and $d(\omega) \approx d_T(\omega)$. By exciting longitudinal and transverse wave packets we examine
individually the energy transport due to longitudinal sound waves and transverse sound waves. Additionally, the random wave
packet allows us to examine how the energy transport separates into contributions
from longitudinal and transverse waves, and we can determine the dominant
energy transport mechanism.

\section{Density of States}
\begin{figure}
\includegraphics[width=0.45\textwidth]{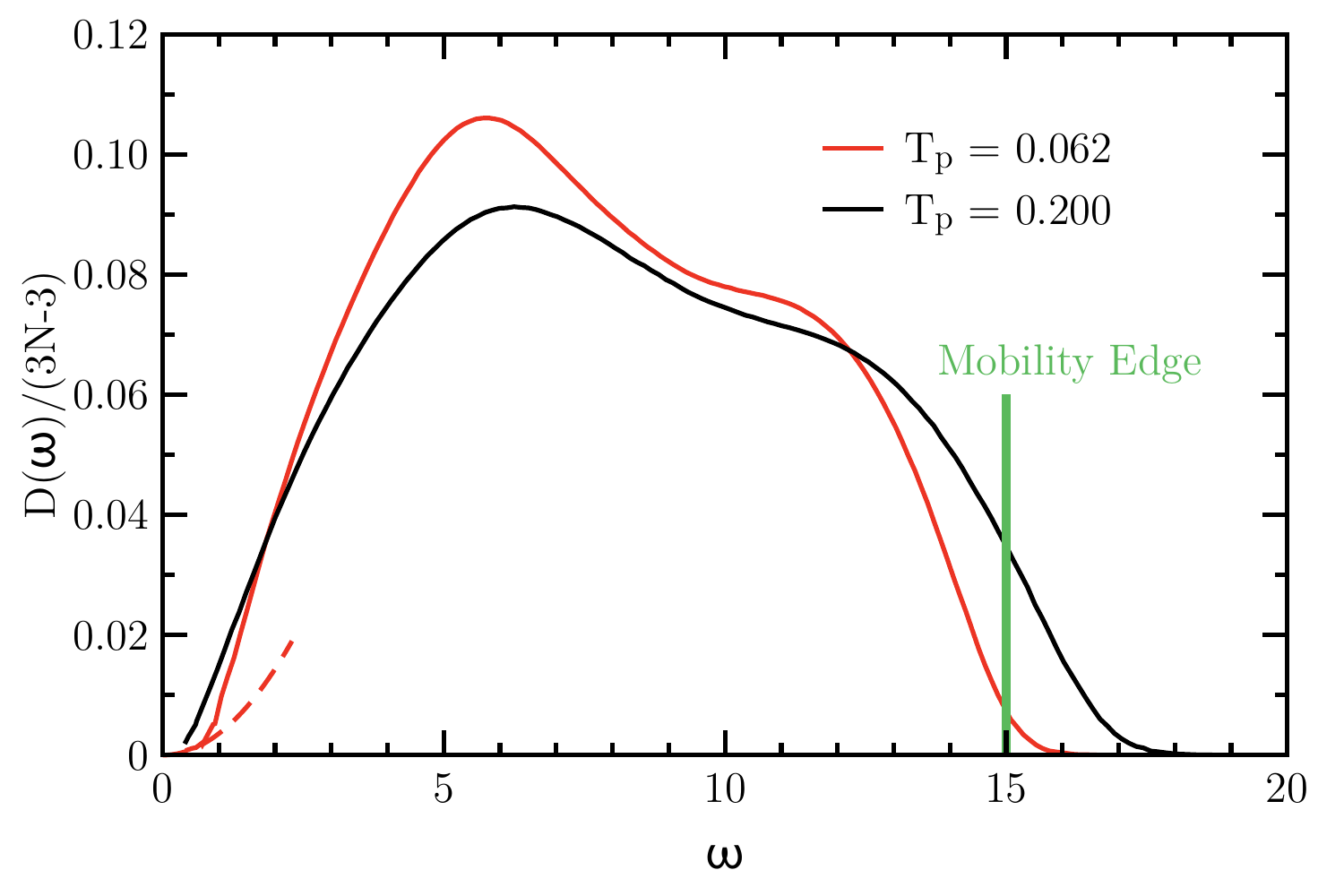}
\caption{\label{dos}Normalized density of states for a stable glass $T_p = 0.062$ (red line) and
for a poorly annealed glass $T_p = 0.2$ (black line). The dashed red line is the Debye prediction
for the density of states for $T_p = 0.062$. The green line marks the mobility edge, where the
energy remains localized and does not propagate.}
\end{figure}

Before we discuss the temperature and stability dependence of the thermal conductivity, we first
examine the density of states. Shown in Fig.~\ref{dos} is the density of states $D(\omega)/(3N-3)$
normalized so that its integral over all
frequencies is equal to one, and thus does not depend on the system size.
We show the density of states for a stable glass with parent temperature $T_p = 0.062$ (red line)
and a  poorly annealed glass with $T_p = 0.2$ (black line).
The dashed red line is the Debye density of states $3 \omega^2/\omega_D^3$ for $T_p = 0.062$,
where $\omega_D^3 = (18 \pi^2 \rho)/(v_L^{-3}+2v_T^{-3})$,
$v_L$ is the longitudinal speed of sound, and $v_T$ is the transverse speed of sound.
The speed of sound was obtained from our previous work on sound attenuation \cite{Wang2019}.
The excess modes above
the Debye prediction for the low-frequency modes are clearly visible. In previous works, it was shown that the
low frequency density of states can be divided into contributions due to extended and quasi-localized modes \cite{Lerner2016,Wang2019,Mizuno2017}.
The density of states of the extended modes agrees with the Debye prediction,
while the density of states of the low-frequency localized modes
scales as $\omega^4$. This scaling of the localized modes has been observed in several simulational studies
\cite{Wang2019,Lerner2016,Mizuno2017,Lerner2017,Kapteijns2018,Angelani2018} and predicted
using different theoretical arguments
\cite{Buchenau1991,Bucenau1992,Schober1996,Gurevich2003,Schirmacher2007,Benetti2018,Stanifer2018,HIkeda2019}.

The more stable glass has fewer low frequency modes, which can be attributed to an increase in $\omega_D$, which
is mainly driven by an increase in the shear modulus (the transverse sound speed). However, there is also a decrease in the
number of quasi-localized modes \cite{Wang2019}. The poorly annealed glass has more modes in the high frequency regime,
above $\omega \approx 13$. The contribution of these modes
to the thermal conductivity will depend on whether the modes are diffusive or localized, since the localized modes
do not contribute to the thermal conductivity in the harmonic approximation.
For frequencies between $\omega \approx 1.7$ and $\omega \approx 12.3$
the density of states is greater for the stable glass.

\section{Energy Transport}
\label{energyT}

Here we will examine the energy transport that follows exciting a longitudinal wave packet, a transverse wave packet, and a random wave
packet. At low frequencies, after exciting the longitudinal wave packet the energy moves \textit{via} longitudinal sound waves and after
exciting the transverse wave packet the energy moves \textit{via} transverse sound waves.  At high frequencies, when
the energy transport is diffusive, the two excitations produce the same results. However, the situation is different for the
random wave packet. Here we will find that at low frequencies the energy transport divides itself into a longitudinal and transverse
contribution that travel ballistically at the speed of longitudinal and transverse sound, respectively.

\begin{figure}
\includegraphics[width=0.45\textwidth]{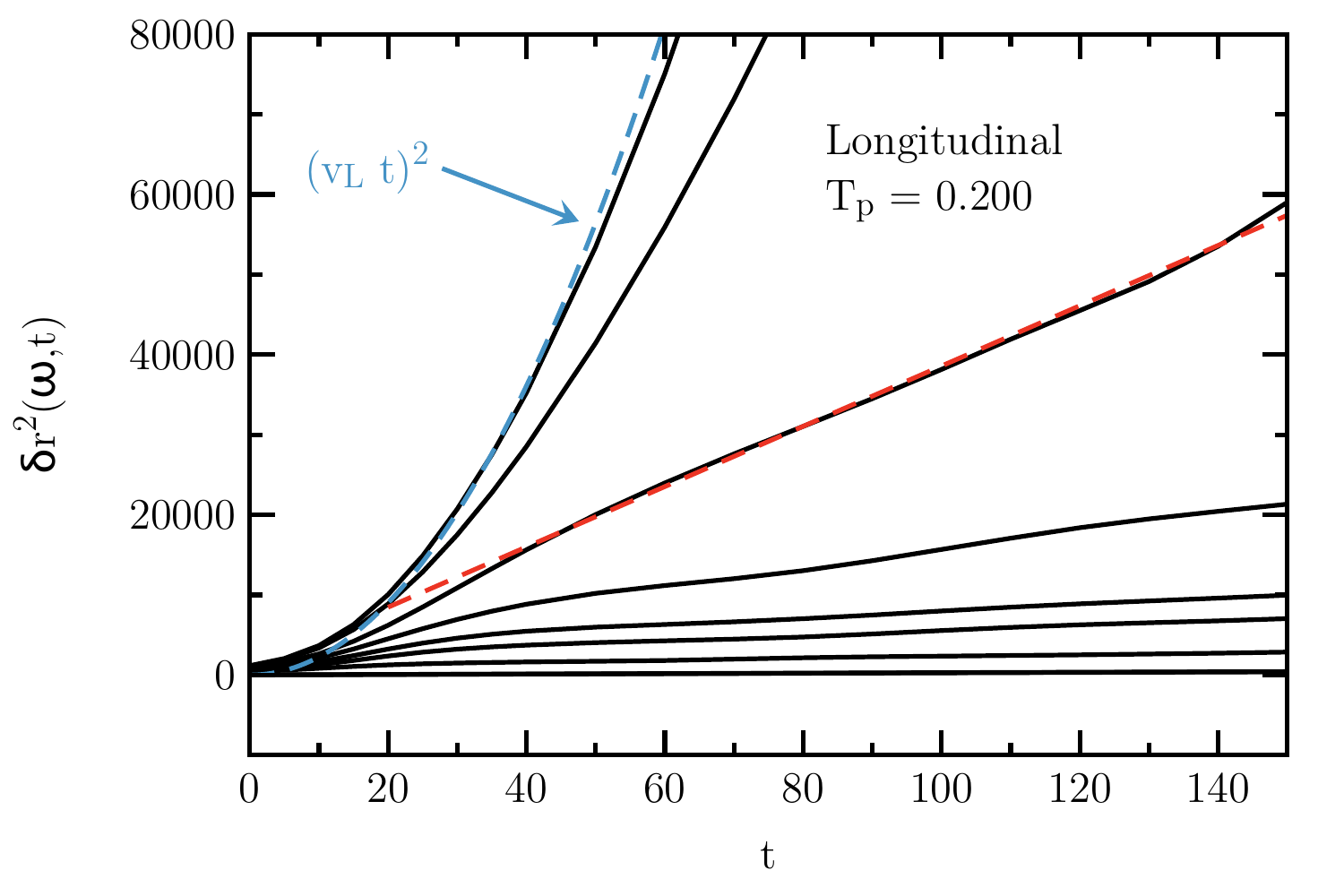}
\caption{\label{Tp2dr2}$\delta r^2(\omega,t)$ for $\omega = 0.3$, 0.4, 0.6, 0.8, 1.0, 1.2, 1.4, 5.0, and 10.0 listed from top to bottom.
The red dashed line is a fit to $\omega = 0.6$ for $t \ge 50$. The blue dashed line is $[v_L(\omega) t]^2$
for $\omega = 0.3$.}
\end{figure}

In Fig.~\ref{Tp2dr2} we show examples of the time dependence of the mean square width of the wave packet,
$\delta r^2(\omega,t)$, for
$T_p = 0.2$ for $\omega = 0.3$, 0.4, 0.6, 0.8, 1.0, 1.2, 1.4, 5.0, and 10.0 listed from top to bottom. The external force starts
at $t=-50$, reaches its maximum at $t=0$ and is effectively zero by $t=50$. If the energy transport is diffusive, we can fit
$\delta r^2(\omega,t) = 2 d(\omega) t + a$ to obtain the diffusivity $d(\omega)$. We show such a fit to $\omega = 0.6$ as a red dashed line.
Once the phonons are the main carriers of the energy, then the energy transport is ballistic and $\delta r^2(\omega,t) = (v_L(\omega) t)^2$,
which is shown as the blue dashed line. The longitudinal speed of sound $v_L(\omega)$ obtained from previous
work on sound attenuation \cite{Wang2019a}.
We observe that the calculated $\delta r^2(\omega,t)$ nearly matches this prediction for $\omega = 0.3$.

\begin{figure}
\includegraphics[width=0.45\textwidth]{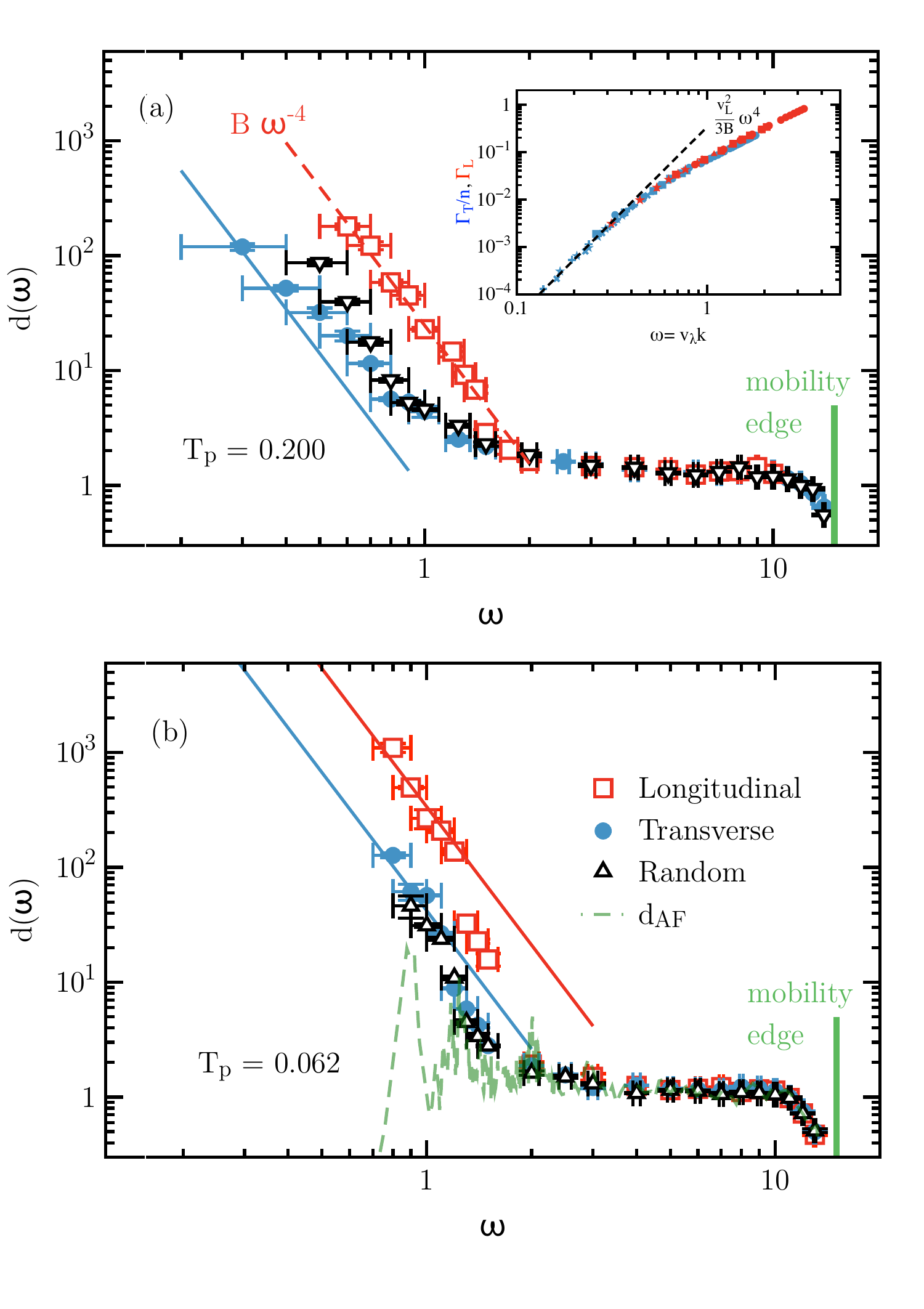}
\caption{\label{diff}Energy diffusivity for $T_p = 0.2$ (a) and $T_p = 0.062$ (b). The red squares are
results for the longitudinal excitation, the blue circles are results for the transverse excitation, and the black
triangles are results for the random excitation. The dashed red line in (a) is a fit to $d(\omega) = B \omega^{-4}$.
We use the phonon gas
model and calculations of sound attenuation to predict $d(\omega)$ at low frequencies, and these predictions
are given by the solid blue (transverse) and red (longitudinal) lines.}
\end{figure}

In Fig.~\ref{diff} we present the thermal diffusivity
$d(\omega)$ calculated for longitudinal (red squares), transverse (blue circles), and random (black triangles)
excitations for a poorly annealed glass, $T_p = 2.0$ (a), and a stable glass, $T_p = 0.062$ (b). At high frequencies, $\omega > 15$,
the energy stays localized. This limit is denoted as the mobility edge in Fig.~\ref{diff} and Fig.~\ref{dos}. The mobility
edge is not sensitive to the glass's stability.  However, as seen in Fig.~\ref{dos} there are many more vibrational modes above the mobility edge
for the poorly annealed glass than for the stable glass.

With decreasing frequency, the diffusivity increases between $\omega \approx 15$ until $\omega \approx 10$.
Between $\omega \approx 10$ and $\omega \approx 2$ the diffusivity is nearly constant and independent of the
nature of the excitation.  As can be seen from Fig.~\ref{dos}, there are more modes within the plateau of
the diffusivity for the stable glass than for
the poorly annealed glass. We also show the diffusivity calculated using the method of Allen and Feldman \cite{Allen1989, Allen1993}
$d_{AF}(\omega)$ (dashed green line) for $T_p = 0.062$
and find excellent agreement within the plateau region, which verifies the physical picture of $d_{AF}$.
Due to the small size of the simulation box, it is not possible to obtain diffusivity at small $\omega$ using the method of
Allen and Feldman. Additionally, it has been recently observed that there are finite size effects in calculations of vibrational modes from the
Hessian matrix \cite{Wang2019a,Lerner2019JCP, Bouchbinder2018}, which may result in calculated thermal diffusivity $d_{AF}(\omega)$
that does not correspond to the thermodynamic limit.

Below $\omega \approx 2$ the diffusivity rapidly increases with decreasing $\omega$ for each type of
excitation. The departure from the plateau is independent of the glass's stability.  Therefore, the
contribution to the thermal conductivity due to the plateau region is due to the difference in the
density of states and not due to the thermal diffusivity, since the thermal diffusivity is stability independent
over this region.
The longitudinal diffusivity increases faster with decreasing $\omega$ than both the transverse and
the random excitation. For low frequencies, both the longitudinal and the transverse diffusivities appear to increase as $\omega^{-4}$
as shown by the red and blue lines in Fig.~\ref{diff}.

For our poorly annealed glass, we fit $d(\omega)$ for the longitudinal excitation for $\omega < 1.5$ to $d(\omega) = B \omega^{-4}$, which
is shown as the red dashed line in Fig.~\ref{diff}. We note that for $\omega > 0.6$ we clearly observe
a linear increase of $\delta r^2(\omega,t)$ with time $t$, as shown in Fig.~\ref{Tp2dr2}. This linear time dependence
is expected for diffusive energy transport, but in Section \ref{density} we will see that for a range of frequencies
the energy density is not well described by a Gaussian distribution indicative of diffusive energy
transport.

We now show that the frequency dependence of $d(\omega)$ is consistent with $d(\omega) = v_L \ell(\omega)/3$ where
the mean free path $\ell(\omega) = v_L/\Gamma_L(\omega)$ and $\Gamma_L(\omega)$
is the sound attenuation coefficient. In earlier work we found that transverse sound attenuation
$\Gamma_T(\omega)$ could be rescaled by a constant factor so that it overlaps with longitudinal sound attenuation
$\Gamma_L(\omega)$\cite{Wang2019a}. In the inset to Fig.~\ref{diff}(a) we show this rescaling and the dashed line
shows $\Gamma_L(\omega) = [v_L^2/(3 B)] \omega^4$ where $B$ is obtained from the fit to $d(\omega)$ shown in the main
plot in Fig.~\ref{diff}(a). The scaling of $d(\omega)$ smoothly continues into the propagating regime where it is no longer
appropriate to consider the energy transport as diffusive.

In the previous paragraph we showed the the low frequency behavior of thermal diffusivity calculated for the
longitudinal excitation can be used to reproduce the low frequency behavior of the sound attenuation. Now, we will
show that the opposite can also be done. To this end we used the low frequency behavior of $\Gamma_\lambda(\omega)$
for longitudinal sound waves ($\lambda = L$) and for transverse sound waves ($\lambda = T$) obtained from
previous work \cite{Wang2019a} to predict the low frequency behavior of $d(\omega)$. The results are shown as
solid lines in Fig.~\ref{diff}(a) and (b). From the low frequency behavior $\Gamma_\lambda(\omega) = A_\lambda \omega^{4}$
we predict that the thermal diffusivity should be given by $v_L^2/(3 A_L \omega^4)$ and $2 v_T^2/(3 A_T \omega^4)$,
for the longitudinal and transverse excitation, respectively. The factor of
2 for the transverse excitation is due to the two polarizations.
The red line illustrates the predicted behavior of the thermal diffusivity for
the longitudinal excitation and the blue lines show the predicted behavior of the thermal diffusivity for
the transverse excitation.
We observe this smooth continuation of the
diffusivity from diffusive energy transport to ballistic energy transport with decreasing frequency for each type of excitation.

It is natural to assume that $d(\omega) = v \ell(\omega)/3$ up until the sound waves are no longer well defined, which
is generally associated with the Ioffe-Regel limit. We determined the Ioffe-Regel limit for
this system for both the longitudinal and transverse sound waves. The Ioffe-Regel limit $\omega_{IR}$ for transverse sound, which is
lower than for longitudinal sound,
for $T_p = 0.062$ is $\omega_{IR} = 1.74$ and for $T_p = 0.200$ it is $\omega_{IR} = 0.9$. The $\omega^{-4}$ scaling does not extend to these
frequencies for the transverse sound waves for either parent temperature, and thus the Ioffe-Regel criteria does
not determine the cutoff for the $\omega^{-4}$ energy transport.  However, it does give an upper bound.

For most frequencies the random excitation follows the transverse excitation, but we observe statistically significant deviations from this
behavior for $T_p = 0.200$ at the smaller frequencies. To get some insight into these deviations, in the next section we will examine in detail the
time dependence of the energy density.

\section{Energy Density}
\label{density}

The physical interpretation of the energy diffusivity $d(\omega)$ proposed by by Allen and Feldman \cite{Allen1989,Allen1993}
is based on the thought experiment that considers the time evolution of a wave packet
narrowly peaked at $\omega$ and initially spatially localized. The square of the width of the wave packet at a time $t$ divided
by $2 t$ is $d(\omega)$ \cite{Allen1993}. This is the operational definition we used in Section
\ref{energyT}. However, the wave packet does not always propagate diffusively and for low frequencies the square of the width increases as $t^2$.
In this section we examine time dependent energy density $E(\omega,x,t)$ for a random excitation.
We compare this energy density with those resulting from the transverse and longitudinal excitations. In this way we clarify the
role of sound waves in the energy transport. We find that diffusive energy transport
describes the excitation in the plateau region, and very clear wave packets propagating at a constant velocity
emerge for low frequency excitations. However, at intermediate frequencies the wave packets are no longer well described as
diffusive or propagating. Similar sort of behavior has been observed in simulations of amorphous silicon \cite{Beltukov2018,Beltukov2016}.

\begin{figure}
\includegraphics[width=0.45\textwidth]{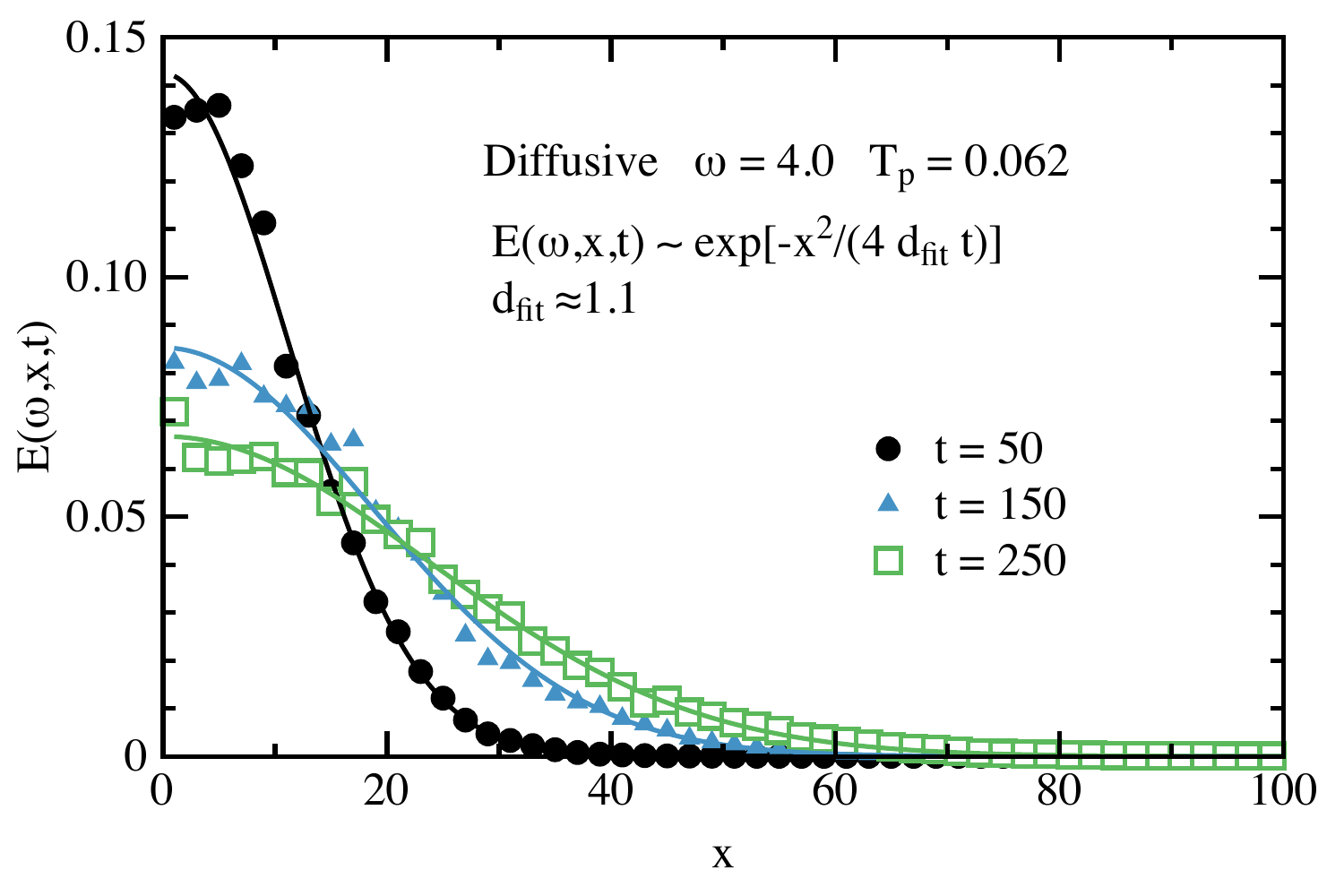}
\caption{\label{diffusion}The energy density for a random excitation for the most stable glass, $T_p = 0.062$, at $\omega = 4.0$.
The solid lines are fits to diffusive energy transport.   }
\end{figure}

We begin by examining $E(\omega,x,t)$ for the random excitation at $\omega = 4$, which is shown in Fig.~\ref{diffusion}  for
$t=50$ (black circles), 150 (blue triangles), and 250 (green squares).
We also show Gaussian fits to the energy density, $E(\omega,x,t) \sim \exp[-x^2/(4 d_{\mathrm{fit}} t)]$ (solid lines),
which describe $E(\omega,x,t)$ well.  We find that $d_{\mathrm{fit}} \approx 1.1$ agrees
well with the value of $1.08$ we found by fitting $\delta r^2(\omega,t)$. There is some ambiguity as to when to
define $t=0$ for the wave packet, which effects the value of $d_{\mathrm{fit}}$. Here it is defined as the time when the force is the largest.

\begin{figure}
\includegraphics[width=0.45\textwidth]{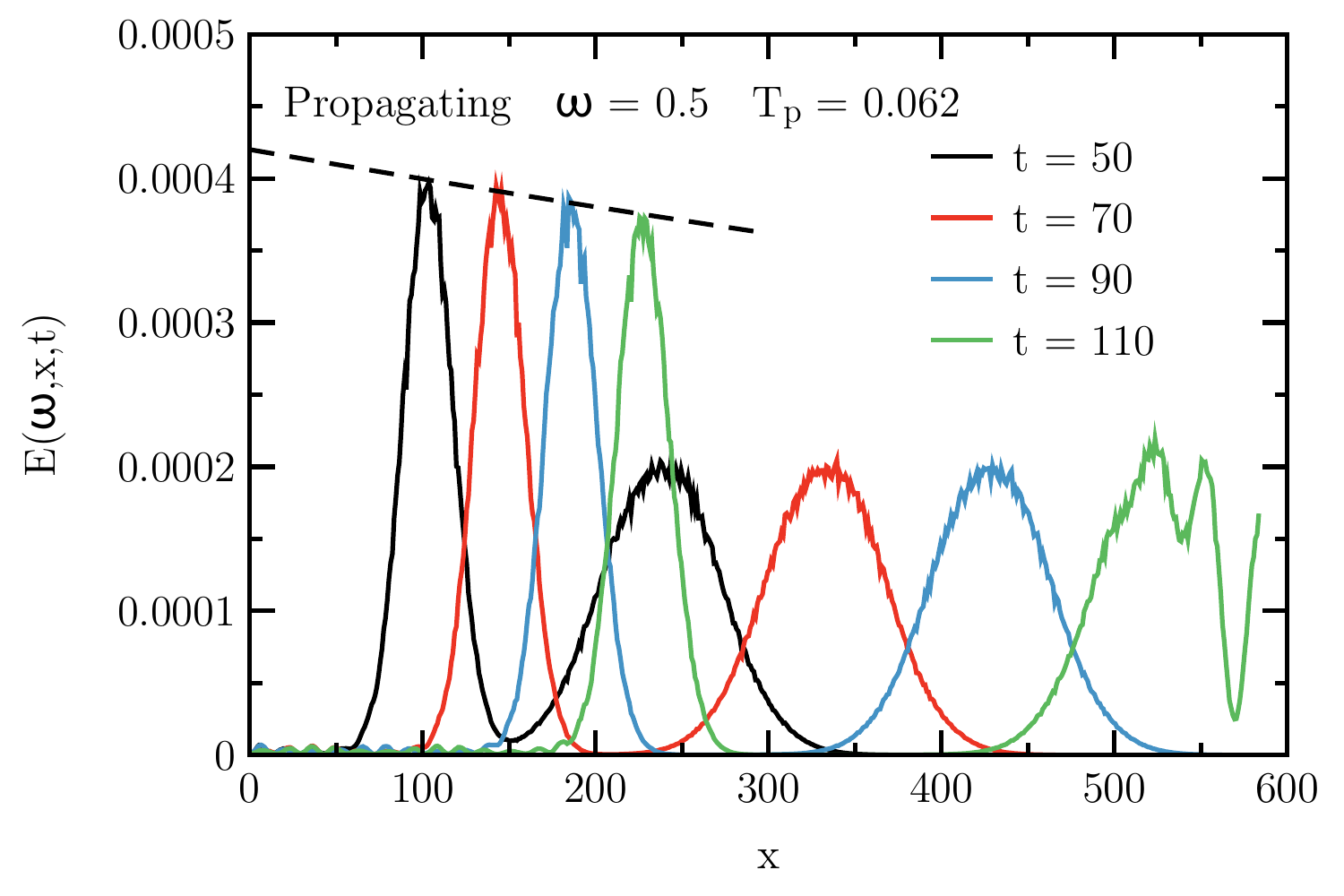}
\caption{\label{prop}The energy density for a random excitation for the most stable glass, $T_p = 0.062$, at $\omega = 0.5$.
The energy breaks up into a propagating transverse wave packet and a propagating longitudinal wave packet.}
\end{figure}

At low frequencies, the time dependence of the wave packet
for a random excitation is very different from that at high frequencies. Shown in Fig.~\ref{prop} are results for
$\omega = 0.5$ for $T_p = 0.062$ for $t=50$ (black),
70 (red), 90 (blue), and 110 (green). The wave packet breaks up into two parts, where one corresponds to energy transport due to longitudinal
sound waves and the other corresponds to energy transport due to transverse sound waves. The longitudinal sound waves travel
faster, and thus the longitudinal part separates from the transverse part and two clear wave packets emerge. For $t=110$ we can see
where the longitudinal wave packet
crosses the boundary of the simulation box and interacts with itself. To confirm this interpretation of the two wave packets,
we verified that the center of the transverse and longitudinal wave packet moves at a velocity $v_T(\omega)$ and
$v_L(\omega)$, respectively.

The mean free path of the transverse excitation $\ell(\omega)$ is given by $\ell(\omega) = 3 d(\omega)/v_T(\omega)$,
and the relationship between $d(\omega)$ and sound attenuation $\Gamma(\omega)$ was discussed in
Section \ref{energyT}. In previous work we demonstrated that $\Gamma(k) = B_T k^4$ for
small $k$, and thus $\ell(\omega) = 2 v_T^6/(B_T \omega^4)$ assuming a linear dispersion relation $\omega = v_T k$.
For $\omega = 0.5$ we use our fit to $\Gamma(k)$ from the previous work \cite{Wang2019a} and obtain $\ell(0.5) \approx 2011$.
Shown as a dashed line in Fig.~\ref{prop} is $a \exp[-x/\ell(0.5)]$, and we find that from the
low frequency (small wavevector) sound attenuation one can predict the decay of
the envelope of the transverse wave packet. To determine $\ell(\omega)$ from the decay of the envelope of the wave packet
is difficult due to small decay over the available time range, but is conceptually possible.

\begin{figure}
\includegraphics[width=0.45\textwidth]{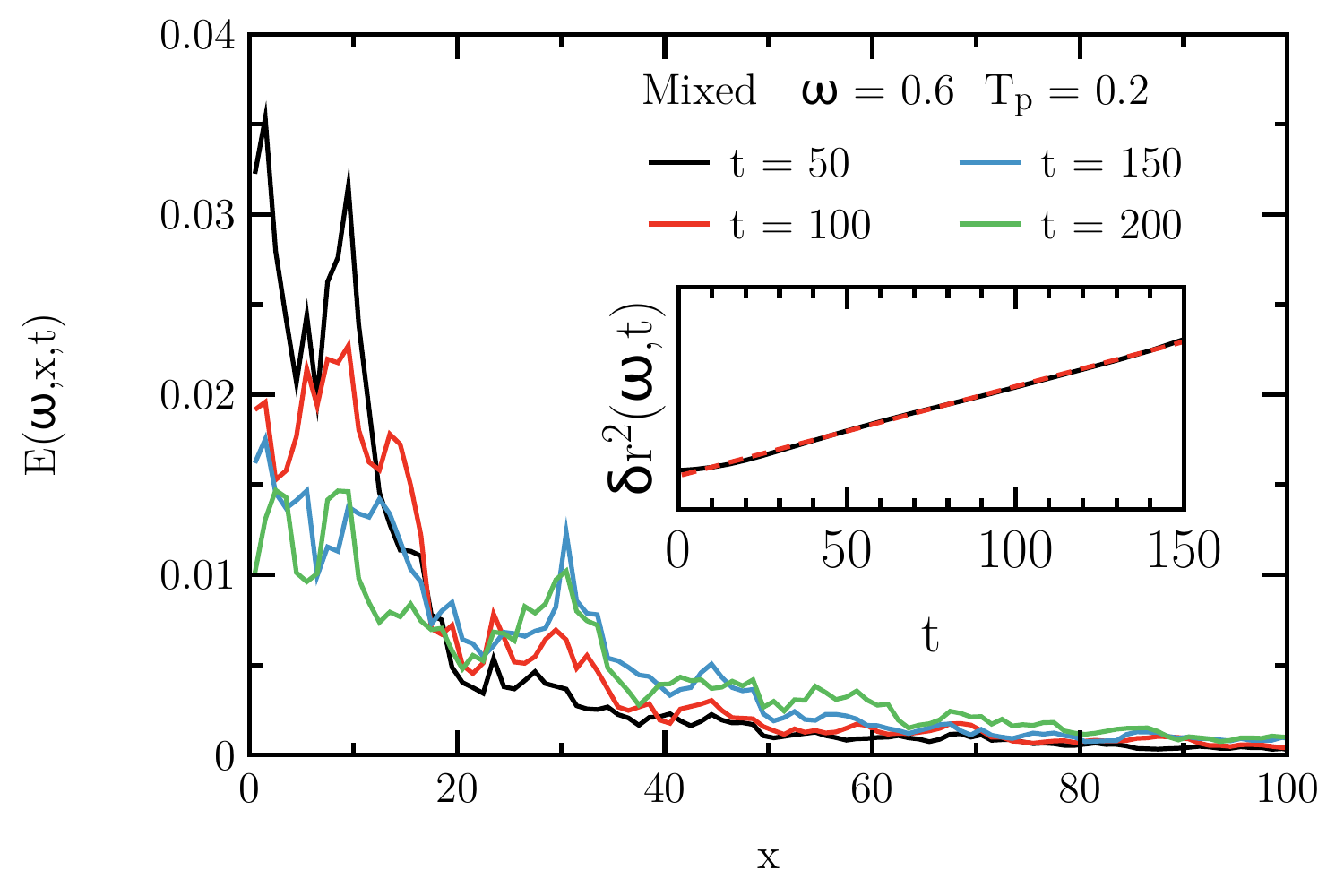}
\caption{\label{mix} The energy density for a random excitation for a poorly annealed glass, $T_p = 0.2$, at $\omega = 0.6$.
The inset shows the spread of the wave packet is linear, but the main figure demonstrates that the energy density is not
Gaussian.}
\end{figure}

For $\omega$ between the pure diffusive regime and the pure ballistic regime the energy density has a very
different time dependence. Shown in Fig.~\ref{mix} is $E(\omega,x,t)$ for $\omega = 0.6$, $T_p = 0.2$
at the time $t=50$ (black), 100 (red), 150 (blue) and 200 (green).
Here we do not observe a Gaussian distribution of the energy density at any time, and there exists a long
tail in the energy density. However, the width characterized by $\delta r^2(\omega,t)$ grows linearly, which
is shown in the inset. This linear growth allows us to calculate $d(\omega)$ for $\delta r^2(\omega,t)$.

\begin{figure}
\includegraphics[width=0.45\textwidth]{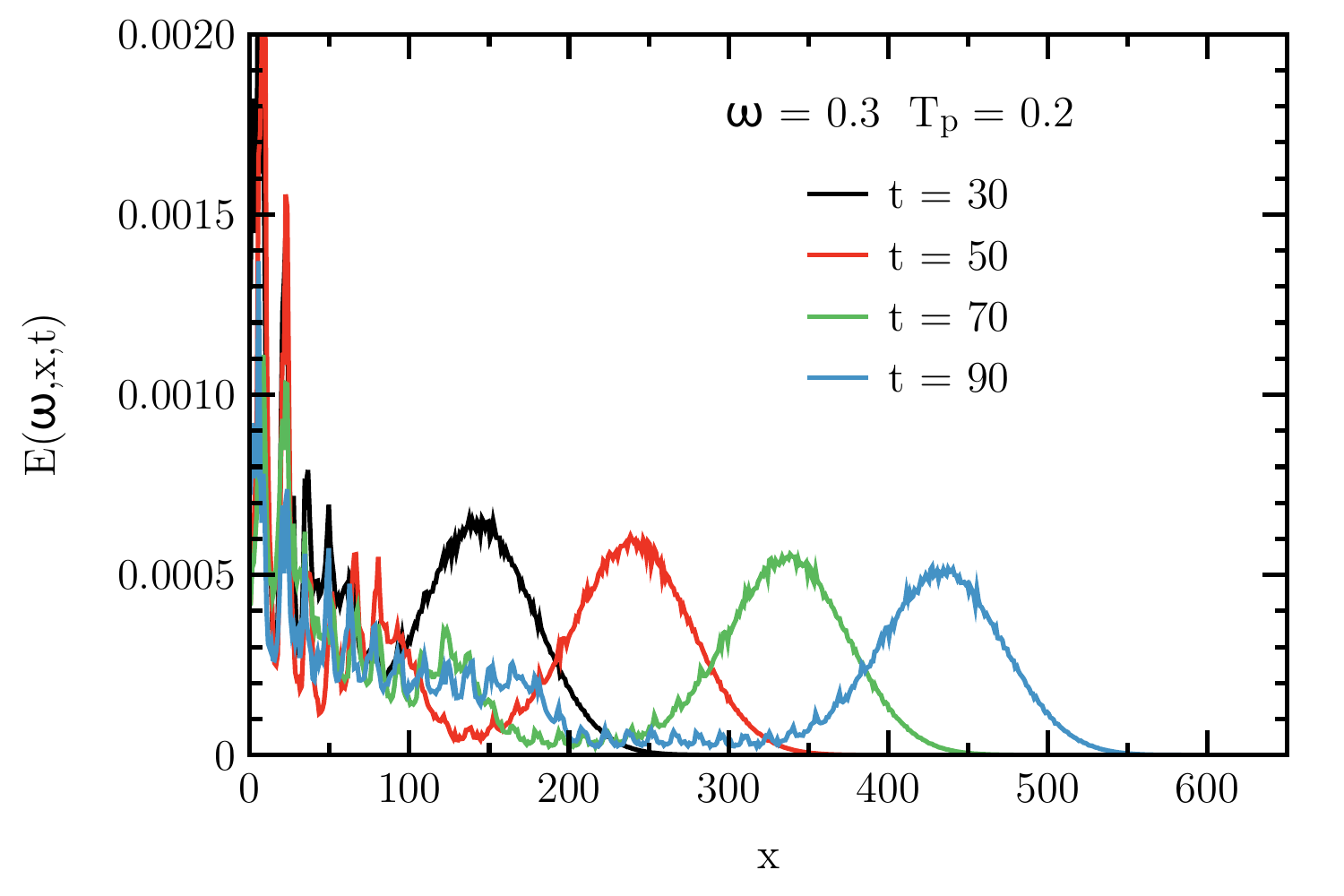}
\caption{\label{propmix} The energy density for a random excitation for a poorly annealed glass, $T_p=0.2$, at
$\omega = 0.3$. There is a propagating wave packet that moves at the constant speed of a longitudinal
wave for $\omega = 0.3$ and a non-propagating part that behaves as a transverse excitation at the same frequency.
}
\end{figure}

Energy transport does not begin to become ballistic for both the longitudinal and transverse
waves for the same frequency for our poorly annealed glass.
This difference in the energy transport of a longitudinal and transverse excitation
for $\omega = 0.3$ and $T_p = 0.2$ is shown in Fig.~\ref{propmix}. Here we observe two different types of behavior.
By comparing with the longitudinal excitation, we find that there is a propagating wave packet that
is proportional to the longitudinal excitation. The propagating part moves at a constant velocity
that is equal to the velocity of the longitudinal sound wave.
The other contribution to the energy density behaves much like the transverse excitation
for this frequency and parent temperature.

The energy transport at $\omega = 0.3$ and $T_p = 0.2$ is carried by a longitudinal sound wave
but not by a transverse sound wave. Therefore, at least for our poorly annealed glass,
there is a narrow frequency window where
longitudinal sound waves significantly contributes to the energy transport, but  transverse sound
waves do not. For low frequencies, the energy transport will be dominated by the transverse sound
waves. We never observed a similar scenario for the stable glass, $T_p = 0.062$, but cannot rule out
that one exists over a narrow frequency range. This difference in the frequency at which energy transport is dominated by the sound
waves for longitudinal and transverse sound results in the difference between $d(\omega)$ given
by a random excitation and the transverse excitation in Fig.~\ref{diff}(a).

\section{Discussion and Conclusions}

We found that the frequency dependence of the thermal diffusivity in glasses, as calculated within the harmonic approximation,
can be divided into four main regions. For low $\omega$ energy transport is
dominated by transverse sound waves whose attenuation $\Gamma$ obeys a Rayleigh scattering law \cite{Wang2019a,Mizuno2018}.
Therefore, at
low frequencies $d(\omega) = A_{\mathrm{low}} \omega^{-4}$ where $A_{\mathrm{low}}$ can be predicted from the
attenuation of transverse sound waves. There is an intermediate regime where the longitudinal sound waves dominate
the energy transport, but this regime may be very narrow or not exist for well annealed glasses. At high frequencies the
diffusivity is nearly flat up to a cutoff frequency $\omega_m$. The transition between the flat diffusivity and the
asymptotic $\omega^{-4}$ scaling occurs over a range $\omega_1 < \omega < \omega_2$ where
$\omega_1$ is stability dependent but $\omega_2$ only weakly depends on stability if at all. We find that
the Ioffe-Regel frequency is within that window, but does not mark the upper end or the lower end of the frequencies.
The lower end of the transition region, $\omega_1$, is stability dependent, but the upper end $\omega_2$ is independent
of the stability.

\begin{figure}
\includegraphics[width=0.45\textwidth]{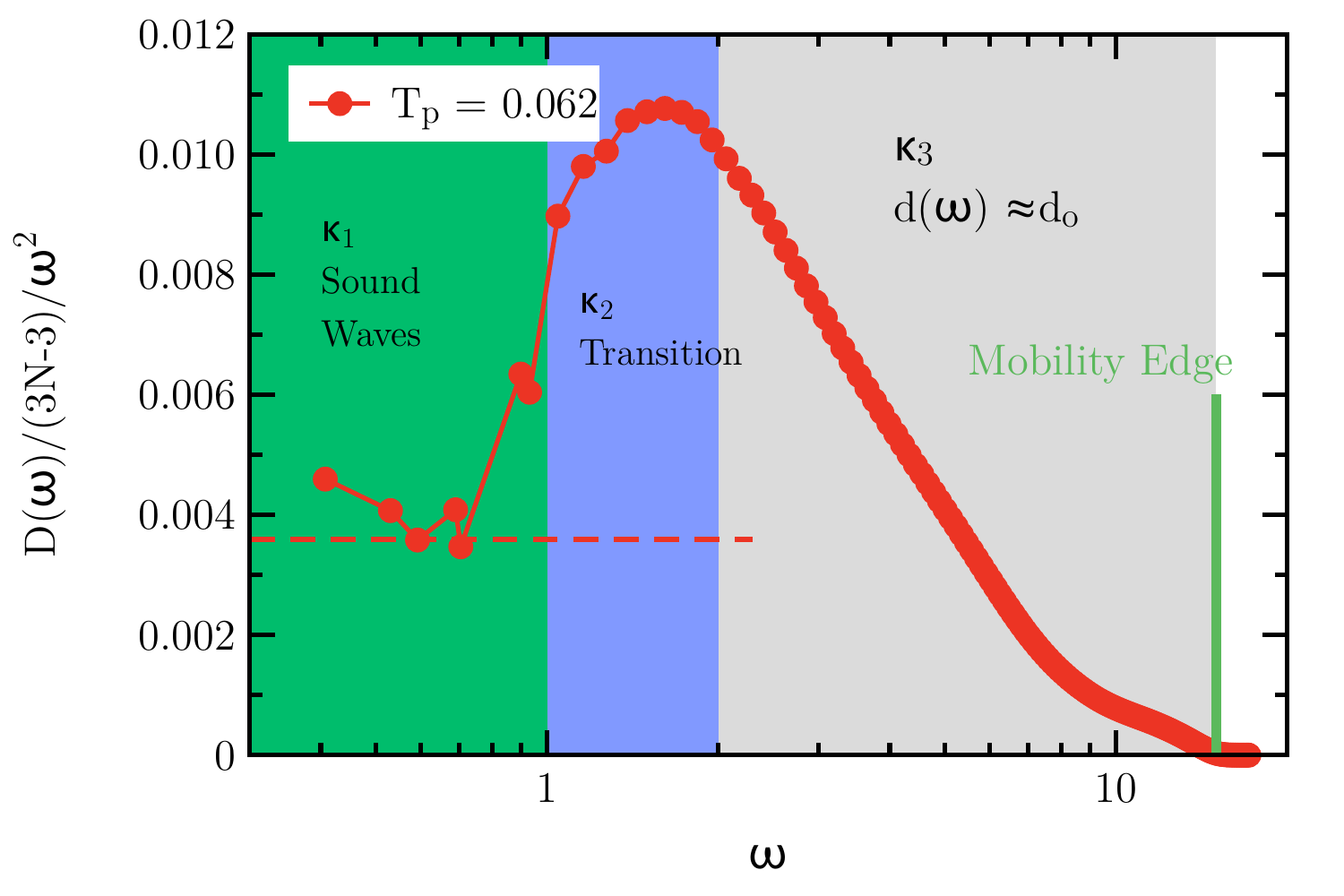}
\caption{\label{rdos}The reduced density of states for $T_p = 0.062$ with different regions of energy transport highlighted.
The green region indicates the frequencies where sound waves dominate the energy transport, and the gray region is the
region of nearly constant diffusivity. The blue region indicates the transition region between the two. }
\end{figure}

These observations lead to  breaking the harmonic approximation to the thermal conductivity into three main contributions
$\kappa \approx \kappa_1 + \kappa_2 + \kappa_3$ where $\kappa_1$ is the contribution from sound waves, $\kappa_2$ is the
contribution from the transition region, and $\kappa_3$ is the contribution from the nearly constant region of diffusivity.
Shown in Fig.~\ref{rdos} is the reduced density of states $D(\omega)/(3N-3)/\omega^2$
for $T_p = 0.062$ with three regions highlighted. For $\omega < 1.0$ sound waves are predominantly responsible
for energy transport, and this is indicated by the green region. For $1.0 < \omega < 2.0$ there is a change
to a nearly flat $d(\omega) \approx d_0$, and this transition region is highlighted  light blue. For $\omega > 2.0$ the
diffusivity $d(\omega) \approx d_o$ is nearly constant up until the mobility edge at $\omega \approx 15$. The region
of nearly constant $d(\omega)$ is highlighted gray.

At low temperatures the $\kappa_1$ term would dominate due to the weight $C(\omega,T)$. It has been established
that the low frequency density of states can be divided into two parts. One part is due to extended modes, which obey
Debye scaling, and one part is due to low-frequency, quasi-localized modes \cite{Wang2019,Mizuno2017}. For these low frequencies,
$D(\omega)/(3N-3) = 3 \omega^2/\omega_D^3 + A_4 \omega^4$ where $\omega_D^3 = [(18 \pi^2 \rho)/(v_L^{-3} + 2 v_T^{-3}]$.
The value of $A_4$ represents the contribution to the density of states from the
quasi-localized modes, and thus is not the main contribution to the energy transport. Neglecting the contribution due to the
quasi-localized modes we can write $\kappa_1$ as
\begin{equation}
\kappa_1 \approx  \frac{6 \rho \hbar v_T^2}{\omega_D^3 B_T} \frac{1}{T} \int_0^{x_1} dx \frac{e^x}{(e^x-1)^2},
\end{equation}
where $x = \hbar \omega/(k_B T)$ and $B_T$ is the coefficient that describes sound attenuation $\Gamma(\omega)$,
$\Gamma(\omega) = B_T \omega^4$. The integral diverges due to the $x=0$ limit. This divergence can be avoided by
setting a lower limit to the integration at $\omega_{\mathrm{min}} = v_T 2 \pi/L$ where $L$ is the length of the amorphous solid
or the divergence can be avoided by including anharmonic contributions.
Two level states are the most likely candidate for the low frequency anharmonic contribution. We note
that $\kappa_1 \sim 1/T$ and would result in a leveling off of the $T^2$ contribution that arises from two level states with increasing temperature.

The second contribution $\kappa_2$ represents the transition between the low frequency $\omega^4$ contribution
to $d(\omega)$ and the nearly flat diffusivity at higher $\omega$. For our poorly annealed glass
there is a range of $\omega$ where there is a significant contribution to $d(\omega)$ due to
longitudinal sound waves, but no contribution due to transverse sound waves. We do not see this behavior
for our stable glass and $d(\omega)$ drops below the extension of the asymptotic small $\omega$
sound wave result.
This intermediate regime extends over a limited range and deserves more study. We determined
the frequency of the boson peak $\omega_{BP} = 1.63$ for $T_p = 0.062$ and $\omega_{BP} = 0.713$
for $T_p = 0.2$, which is in this transition region of $d(\omega)$.

\begin{figure}
\includegraphics[width=0.45\textwidth]{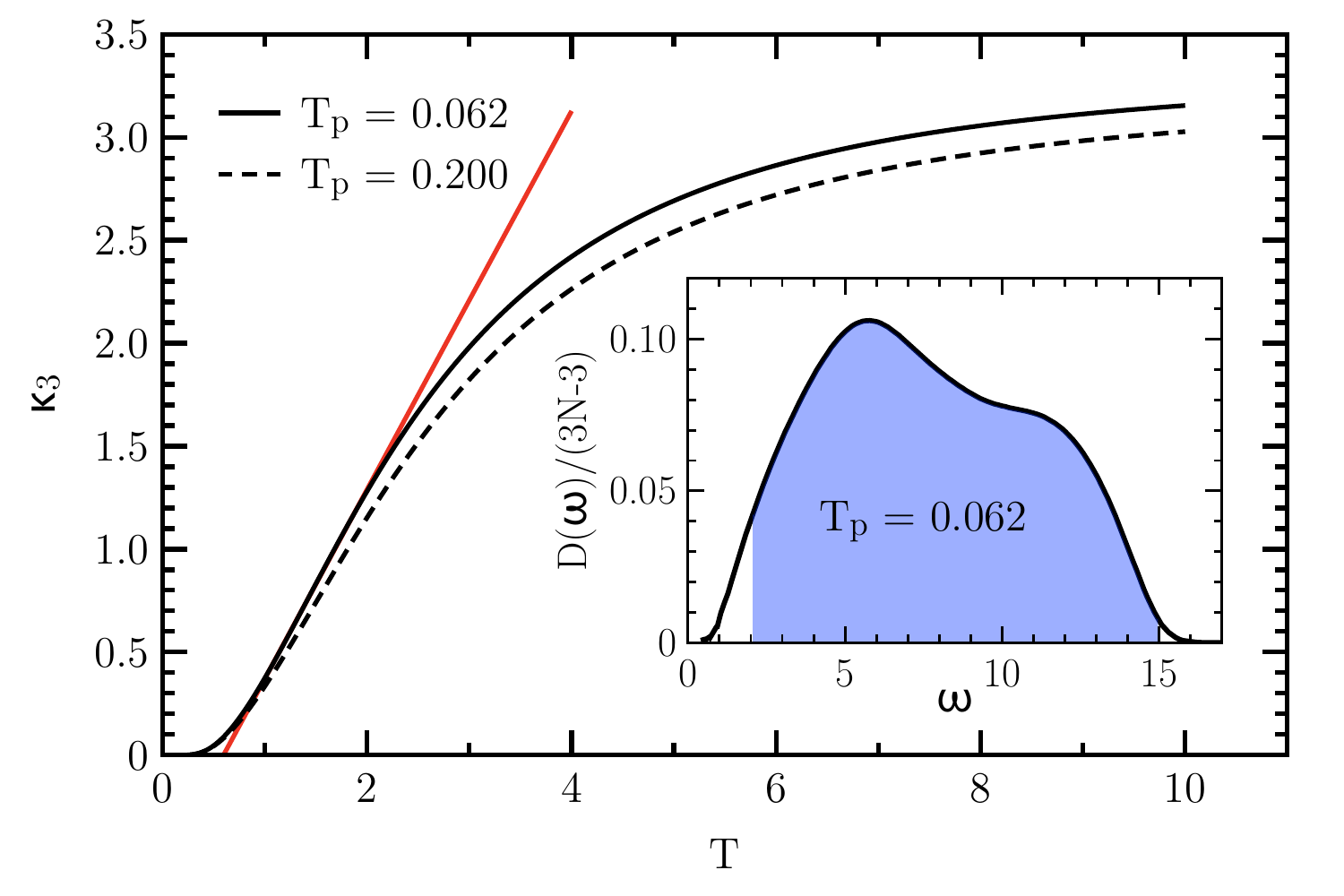}
\caption{\label{dos1}The contribution to the thermal conductivity due to
the region of nearly constant diffusivity, $\kappa_3$. The red line is a fit
to $\kappa_3$ between $T=1$ and $T=2$. (inset) The vibrational density of states for $T_p = 0.062$ with the shaded region
showing the range of $\omega$ of approximate
constant diffusivity.}
\end{figure}

The third contribution $\kappa_3$ is due to the region of approximately flat diffusivity.
If we assume that $d(\omega) \approx d_0$, then
\begin{equation}
\kappa_3 \approx 3 \rho d_0 \int_{\omega_2}^{\omega_m} d\omega D(\omega) k_B \left( \frac{\hbar \omega}{k_B T} \right)^2
\frac{e^{\hbar \omega/(k_B T)}}{(e^{\hbar \omega/(k_B T)} - 1)^2}.
\end{equation}
In the inset to Fig.~\ref{dos1} we show this integration range as the shaded area under the density of states for
$T_p = 0.062$. Most of the vibrational
modes are within this region, with 94\% of the vibrational density of states within this region
for $T_p = 0.2$ and 97\% within this region for $T_p = 0.062$.
Unlike upon the approach to the unjamming transition \cite{Xu2009}, the density of states does
not demonstrate any regions where both the diffusivity and the density of states are nearly flat. To check the
temperature dependence of this contribution, we numerically integrate $\kappa_3$ assuming that $d(\omega) = d_0 = 1.15$
and set $\hbar$ and $k_B$ to one.

Shown in Fig.~\ref{dos1} is $\kappa_3$ versus temperature for the stable glass $T_p = 0.062$ (solid line) and
the poorly annealed glass $T_p = 0.2$ (dashed line). This contribution to the thermal conductivity is
what is observed for amorphous solids above the $T \approx 10K$ plateau. For low temperatures
$\kappa_3$ is negligible, which corresponds to temperatures at and below the plateau in the thermal
conductivity. After this plateau, and these vibrational states become populated, there is a near linear increase
of $\kappa_3$. The red line is a linear fit to $\kappa_3$ for $1 \le T \le 3$ for $T_p = 0.062$. At high temperatures, $C(\omega,T) \approx k_B$
and $\kappa_3$ saturates.  As suggested by studies \cite{Xu2009} approaching the unjamming transition, the region of
flat diffusivity can accurately describe the behavior of the thermal conductivity above the plateau at approximately $10K$.

More work is needed in order to understand energy transport beyond the harmonic approximation. Two level tunneling
states are postulated to describe many aspects of the universal low temperature properties of amorphous solids
\cite{Zeller1971,Anderson1972,Pohl2002}.
Since these states arise from two nearby energy minima  due to small rearrangements of particles, it
may be possible to identify the classical analogues of these states in classical model glassy systems. This would allow for a more detailed
investigation of the low temperature thermal conductivity. It is also possible to examine anharmonic effects of energy
transport using wave packets by running molecular dynamics simulations using the full potential instead of
the harmonic approximation used here.

\section{Acknowledgements}
L.W., E.F., and G.S. acknowledge funding from NSF DMR-1608086.

\end{document}